%%%%%%%%%%%%%%%%%%%%%%%%%%%%%%%%%%%%%%%%%%%%%%%%%%%%%%%%%%%%%%%%%%%%%%%%%
%%			    Tex instructions			       %%
%%%%%%%%%%%%%%%%%%%%%%%%%%%%%%%%%%%%%%%%%%%%%%%%%%%%%%%%%%%%%%%%%%%%%%%%%
%%%%%%%%%%%%%%%%%%%%%%%%%%%%%% FONT MACROS %%%%%%%%%%%%%%%%%%%%%%%%%%%%%%

\font\titolino=cmbx10
\font\tsnorm=cmr10
\font\tscors=cmti10
\font\tsnote=cmr9

\font\tsnotec=cmti9
%%%%%%%%%%%%%%%%%%%%%%%%%% PRINTING MACROS %%%%%%%%%%%%%%%%%%%%%%%%%%%%%
\magnification=1200
\hsize=148truemm
\hoffset=10truemm
\parindent 8truemm
\parskip 3 truemm plus 1truemm minus 1truemm
\nopagenumbers
\newcount\notenumber

\def\note{\advance\notenumber by 1 \footnote{$^{\the\notenumber}$}}
\def\beginref{\begingroup\bigskip
\leftline{\titolino References.}
\nobreak\smallskip\noindent}
\def\endref{\par\endgroup}
\def\ref#1#2{\noindent\item{\hbox to 25truept{[#1]\hfill}} #2.\smallskip}
\def\beginsection #1. #2.
{\bigskip
\leftline{\titolino #1. #2.}
\nobreak\medskip\noindent}
\def\beginappendix #1.
{\bigskip
\leftline{\titolino Appendix #1.}
\nobreak\medskip\noindent}
\def\beginack
{\bigskip
\leftline{\titolino Acknowledgments.}
\nobreak\medskip\noindent}
%

%

%%%%%%%%%%%%%%%%%%%%%%%%%%%% TEXT MACROS %%%%%%%%%%%%%%%%%%%%%%%%%%%%%%%

\def\Sc{Sch\-warz\-sch\-ild}
\def\Schr{Sch\-r\"o\-din\-ger}

%%%%%%%%%%%%%%%%%%%%%%%%%%% MATH MACROS %%%%%%%%%%%%%%%%%%%%%%%%%%%%%%%
\def\pg{p_\gamma}
\def\vf{\varphi}
\def\pphi{p_\varphi}
\def\pcsi{p_\xi}
\def\px{p_x}
\def\py{p_y}
\def\pw{p_w}
\def\pz{p_z}
\def\tv{t_{\rm GV}}
\def\sl{\sqrt{\Lambda}}
\def\d{\partial}
\def\ra{\rightarrow}

\def\de{\Delta_{\rm FP}}
\def\arcth{{\rm arcth\,}}

\def\frac#1#2{{{#1}/over{#2}}}
\def\Sh{Shan\-mu\-ga\-dha\-san}
%%%%%%%%%%%%%%%%%%%%%%%% JOURNAL MACROS %%%%%%%%%%%%%%%%%%%%%%%%%%%%%%%

\def\APP#1#2#3{{\tscors Astropart.\ Phys.} {\bf #1}, #2 (#3)}

\def\NPB#1#2#3{{\tscors Nucl.\ Phys.} {\bf B#1}, #2 (#3)}

\def\PLB#1#2#3{{\tscors Phys.\ Lett.} {\bf B#1}, #2 (#3)}

\def\JMP#1#2#3{{\tscors J.\ Math.\ Phys.} {\bf #1}, #2 (#3)}

\def\IJMPA#1#2#3{{\tscors Int.\ J.\ Mod.\ Phys.} {\bf A#1}, #2 (#3)}

%%%%%%%%%%%%%%%%%%%%%%%%%%%%%%%%%%%%%%%%%%%%%%%%%%%%%%%%%%%%%%%%%%%%%%%%%
%%			 End of instructions			       %%
%%%%%%%%%%%%%%%%%%%%%%%%%%%%%%%%%%%%%%%%%%%%%%%%%%%%%%%%%%%%%%%%%%%%%%%%%
%%%%%%%%%%%%%%%%%%%%%%%%%%%%%%%%%%%%%%%%%%%%%%%%%%%%%%%%%%%%%%%%%%%%%%%%%
%%			       Title				       %%
%%%%%%%%%%%%%%%%%%%%%%%%%%%%%%%%%%%%%%%%%%%%%%%%%%%%%%%%%%%%%%%%%%%%%%%%%
\null
\vskip 5truemm
\rightline {SISSA 71/96/A}
\rightline {DFTT 22/96}
\rightline{May 9, 1996}
\vskip 12truemm
\centerline{\titolino TIME GAUGE FIXING AND HILBERT SPACE}
\bigskip
\centerline{\titolino IN QUANTUM STRING COSMOLOGY}
\vskip 12truemm
\centerline{\tsnote MARCO CAVAGLI\`A$^\star$}
\smallskip
\centerline{\tsnotec SISSA - International School for Advanced Studies,}
\smallskip
\centerline{\tsnotec Via Beirut 2-4, I-34014 Trieste, Italy,}
\smallskip
\centerline{\tsnotec and}
\smallskip
\centerline{\tsnotec INFN, Sezione di Torino, Italy.}
\bigskip
\centerline{\tsnote VITTORIO DE ALFARO$^\diamond$}
\smallskip
\centerline{\tsnotec Dipartimento di Fisica Teorica dell'Universit\`a di
Torino,}
\smallskip
\centerline{\tsnotec  Via Giuria 1, I-10125 Torino, Italy,}
\smallskip
\centerline{\tsnotec  and}
\smallskip
\centerline{\tsnotec INFN, Sezione di Torino, Italy.}
\vfill
\centerline{\tsnorm ABSTRACT}
\begingroup\tsnorm\noindent
%\baselineskip=10truemm
%%%%%%%%%%%%%%%%%%%%%%%%%%%%%%%%%%%%%%%%%%%%%%%%%%%%%%%%%%%%%%%%%%%%%%%%%
%%			      abstract				       %%
%%%%%%%%%%%%%%%%%%%%%%%%%%%%%%%%%%%%%%%%%%%%%%%%%%%%%%%%%%%%%%%%%%%%%%%%%
Recently the low-energy effective string theory has been used by
Gasperini and Veneziano to elaborate a very interesting scenario for the
early history of the universe (``birth of the universe as quantum
scattering''). Here we investigate the gauge fixing and the problem of
the definition of a global time parameter for this model, and we obtain
the positive norm Hilbert space of states. 
%%%%%%%%%%%%%%%%%%%%%%%%%%%%%%%%%%%%%%%%%%%%%%%%%%%%%%%%%%%%%%%%%%%%%%%%%
%%			   End of abstract			       %%
%%%%%%%%%%%%%%%%%%%%%%%%%%%%%%%%%%%%%%%%%%%%%%%%%%%%%%%%%%%%%%%%%%%%%%%%%
%%%%%%%%%%%%%%%%%%%%%%%%%%%%%%%%%%%%%%%%%%%%%%%%%%%%%%%%%%%%%%%%%%%%%%%%%
%%			       Address				       %%
%%%%%%%%%%%%%%%%%%%%%%%%%%%%%%%%%%%%%%%%%%%%%%%%%%%%%%%%%%%%%%%%%%%%%%%%%
\bigskip
\leftline{\tsnorm PACS: 04.60.Ds, 98.80.Hw\hfill}
\smallskip
\hrule
\smallskip\noindent
\leftline{$^\star$ E-Mail: CAVAGLIA@SISSA.IT\hfill}
\leftline{$^\diamond$ E-Mail: VDA@TO.INFN.IT\hfill}
\endgroup
\vfill
\eject
\footline{\hfill\folio\hfill}
\pageno=1
%\baselineskip=2\normalbaselineskip
\noindent
%%%%%%%%%%%%%%%%%%%%%%%%%%%%%%%%%%%%%%%%%%%%%%%%%%%%%%%%%%%%%%%%%%%%%%%%%
%%			      text				       %%
%%%%%%%%%%%%%%%%%%%%%%%%%%%%%%%%%%%%%%%%%%%%%%%%%%%%%%%%%%%%%%%%%%%%%%%%%
\beginsection 1. Introduction.
In the customary quantum gravity approach to the origin of the universe
[1], use is made of the Wheeler-DeWitt (WDW) equation whose solutions with
appropriate boundary conditions describe the ``tunneling from nothing''. 

This fundamental approach has assumed a renewed interest since the
classical string cosmology [2] describes the formation of a
Friedmann-Robertson-Walker (FRW) universe with essentially the present
characteristics as evolving from the string perturbative vacuum. The
transition from an initial ``pre-big bang'' phase to the present one is
represented in quantum string cosmology by a scattering and reflection
of the WDW wave function in superspace [3,4]. 

Now the WDW equation has the usual problems of measure and definition of
inner product; in the present case Gasperini and Veneziano (in the
following GV) have surmounted the ambiguities of the differential
representation of operators using the symmetry of the classical action,
which is the right recipe, see e.g.\ [5]. 

In view of the renewed interest in the matter we have given a closer
look at the determination of a hermitian Hamiltonian and a positive norm
Hilbert space.  A Hilbert space is requested in quantum mechanics; the
sole WDW equation gives wave functions, but no inner product.
Establishing a positive norm Hilbert space of states is an exercise in
the gentle art of finding a time gauge such that evolution is described
by a unitary operator. Now, this requires the model to be taken
seriously, since the time determined by fixing the gauge is defined as a
suitable function of the canonical coordinates of the problem under
consideration. This may be thought of as inadequate to the complexity of
gravitational systems, but these are the rules of quantum mechanics
applied to minispaces; there is no alternative. 

The two approaches to gauge fixing are 1) quantisation of the constraint (Dirac method)
followed by gauge fixing, or 2) reduction of the canonical space by
introducing a classical time gauge fixing condition and use of the
constraint, followed by quantisation in the reduced space if the reduced
Hamiltonian is hermitian [6]. With a proper gauge condition the two
methods give coincident results. In this way we obtain the definition of
the norm. 
\beginsection 2. Conventions and Definitions.
We gather here the necessary formulae so that the paper is self --
consistent. The definitions and results are essentially as in [3,4] with a
few changes in notation and normalization.

We start from the usual four dimensional low-energy effective string
action [7]
$$S={1\over 2\lambda_s^2}\int_V d^4x\sqrt{-g}e^{-2\Phi}
\left(R+4\d_\mu\Phi\d^\mu\Phi-\Lambda\right)\,,\eqno(2.1)$$
where $\Phi$ is the dilaton, $\Lambda>0$ is the cosmological constant, and
$\lambda_s$ is the fundamental string-lenght parameter. In (2.1) we use
for the Ricci scalar the conventions of Landau-Lifshits [8].  The metric
is assumed to be spatially homogeneus and isotropic: 
$$ds^2=-N^2(t)dt^2+a^2(t)~\omega^p\otimes\omega^p\,.\eqno(2.2)$$
Here $N$ is the lapse function and the scale factor $a$ is positive by
definition. The $\omega^p$'s are the 1-forms that satisfy the
Maurer-Cartan structure equation
$$d\omega^p={k\over
2}\epsilon_{pqr}~\omega^q\wedge\omega^r\,,\eqno(2.3)$$
where $k=0, \pm 1$. Accordingly the dilaton field is assumed to depend
only on time. Let us define
$$\gamma=\ln a, ~~~~~~~
\varphi=\Phi-{3\over 2}\ln a\,.\eqno(2.4)$$
$\varphi$ is usually called the ``shifted dilaton field'' [3]. It is
advisable to use the Lagrange multiplier
$$ \mu~=~N\,e^{2 \vf}\,. \eqno(2.5)$$
Indeed using (2.2-5) in (2.1) one has [3,4] (here and throughout the 
paper we neglect inessential surface terms)
$$S={V\over 2\lambda_s^2}\int dt\left[
3{\dot\gamma^2\over\mu}-4{\dot\varphi^2\over\mu}+\mu
e^{-4\varphi}\left(6k
e^{-2\gamma}-\Lambda\right)\right]\,,\eqno(2.6)$$
where $V$ is the spatial volume element with $a=1$ and dots represent
differentiation with respect to $t$. In the following we will set
$V/2\lambda_s^2=1$. In canonical form the action (2.6) becomes
$$S=\int dt \left\{\dot\gamma\pg+\dot\varphi
\pphi-{\cal H}\right\} \,,\eqno(2.7)$$
where
$$\pg=6{\dot\gamma\over\mu}\,,~~~~~~~
\pphi=-8{\dot\varphi\over\mu}\,,\eqno(2.8)$$
are respectively the conjugate momenta of $\gamma$ and $\varphi$, and
$${\cal H}=\mu
H=\mu\left[{\pg^2\over 12}-{\pphi^2\over 16}+\Lambda
e^{-4\varphi}-6ke^{-2\gamma}\right]\,.\eqno(2.9)$$
Here $H$ is the generator of time-reparametrizations (gauge
transformations); we will simply call it the ``Hamiltonian'' of the
system. The Lagrange multiplier $\mu$ enforces the constraint
$$H=0\eqno(2.10)$$
which expresses the invariance under time-reparametrization. The gauge
transformations generated by $H$ are
$$\eqalignno{&\delta q_i =\epsilon {{\d H}\over {\d p_i}} 
= \epsilon\bigl[ q_i,H\bigr]_P\,,&\hbox{(2.11a)}\cr\cr
&\delta p_i =-\epsilon {{\d H} \over {\d q_i}} 
= \epsilon\bigl[ p_i,H \bigr]_P\,,&\hbox{(2.11b)}\cr\cr
&\delta l = {{d \epsilon} \over {dt}}\,,&\hbox{(2.11c)}\cr\cr}$$
where $q_i=\{\gamma,\varphi\}$, $p_i=\{\pg,\pphi\}$. Throughout the paper
we will set $k=0$ in (2.9), i.e.\ we will consider only flat spacetimes. 
The case of $k=\pm 1$ will be considered elsewhere. As a warm up exercise,
now we illustrate the procedure on the simple case of null cosmological 
constant, corresponding to the D'Alembert Hamiltonian. 
\beginsection 3. The D'Alembert Case.
The case $\Lambda=0$ corresponds to a string with critical dimension [2].
Taking $k=0$ and $\Lambda=0$ in (2.9) the Hamiltonian becomes
$${\cal H}=\mu H=\mu\left[{\pg^2\over 12}-{\pphi^2\over
16}\right]\,.\eqno(3.1)$$
The finite gauge transformations (2.11) can be integrated
explicitly. The result is
$$\eqalignno{&\gamma=\gamma_0+{\pg\over 6}\tau\,,&\hbox{(3.2a)}\cr
&\pg={\rm constant}\,,&\hbox{(3.2b)}\cr
&\varphi=-{\pphi\over 8}\tau\,,&\hbox{(3.2c)}\cr
&\pphi={\rm constant}\,,&\hbox{(3.2d)}\cr
&\tau=\int_{t_0}^t \mu(t')~dt'\,,~~~~~~\mu(t)>0\,.&\hbox{(3.2e)}\cr}$$
where $\gamma_0$, $\pg$, and $\pphi$ are gauge invariant quantities.
We can define the new variables (action-angle variables)
$$\eqalignno{&\xi=6{\gamma\over\pg}\,,&\hbox{(3.3a)}\cr
&\pcsi={1\over 12}\pg^2\,,&\hbox{(3.3b)}\cr}$$
that will be used later. The Poisson relation of $\xi$ and $\pcsi$ is
$$\bigl[\xi,\pcsi\bigr]_P=1\,.\eqno(3.4)$$
Thus $\{\varphi,\pphi,\xi,\pcsi\}$ form a complete set of canonically
conjugate variables. Note that $\pphi$ and $\pcsi$ are gauge
invariant quantities and $\xi$ tranforms by gauge tranformations as
$$\xi\to\bar\xi=\xi+\tau\,.\eqno(3.5)$$
This is the reason for the interest in $\xi$. Eq.\ (3.5) suggests that 
$\xi$ is a proper variable to fix the gauge and obtain a unitary
evolution in the gauge fixed space (see later).	We call the set
$\{\varphi,\pphi,\xi,\pcsi\}$ ``hybrid'' variables because they
are not the maximal gauge invariant choice of canonical coordinates.
Indeed we can identify a maximal set of gauge invariant canonical
variables (we will refer to them as ``\Sh'' variables, see
[9])
$$\eqalignno{&x=\varphi+{3\over 4}
{\gamma\over\pg}\pphi\,,&\hbox{(3.6a)}\cr
&\px\equiv\pphi\,,&\hbox{(3.6b)}\cr
&y\equiv\xi=6{\gamma\over\pg}\,,&\hbox{(3.6c)}\cr
&\py\equiv H={1\over 12}\pg^2-{1\over 16}\pphi^2\,,&\hbox{(3.6d)}\cr}$$
which is a set of canonically conjugate variables.

The variables $x$ and $\px$ are gauge invariant and thus generate
rigid invariance transformations.  Of course the meaning of gauge
invariant variables is transparent in the case of $x$: it is the
initial value of $\vf$.  These variables and the functions $f(x,\px)$
are the observables:  ``The set of the observables is isomorphic to
the set of functions of the initial data'' [10].

For sake of compleneness, we write the generating function of the
canonical transformation $\{\gamma,\pg;\varphi,\pphi\}$ $\ra$
$\{x,\px;y,\py\}$.
$$F=-{3\gamma^2\over y}+{4\over y}(\varphi-x)^2\,.\eqno(3.7)$$
Each  set, $\{\varphi,\pphi;\xi,\pcsi\}$ or $\{x,\px;y,\py\}$, can be
used in the quantisation program and leads to identical results, both in
the Dirac method (quantise before constraining) and in the reduced
method (constrain before quantising). Let us first quantisze in the
hybrid variables. 
\bigskip
\leftline{\it a) Quantisation in Hybrid Variables}
\medskip
Let us start by the Dirac method.  Wave functions are solutions of the
WDW equation. Now, the gauge has to be fixed [5,6,11] in the
scalar product of solutions of the WDW equation. Let us define the 
scalar product as
$$(\Psi_2,\Psi_1)\,=\,\int d[\alpha]\,\Psi_2^* \delta (\Theta)
\Delta_{FP} \Psi_1\,,\eqno(3.8)$$
where $\Theta(\alpha_i)=0$ is the gauge fixing identity, $\alpha_i$
($i=1,2$) are the canonical coordinates, and $\Delta_{FP}$ is the
Faddeev-Popov (FP) determinant. $d[\alpha]$ is the off-shell measure and
is of course defined in the unconstrained phase space. 

Now the first problem is the choice of the variables and of the measure.
We require the measure to be gauge invariant and invariant with respect
to the rigid symmetries of the system. The choice $d[\alpha]=d\pphi
d\pcsi$ is gauge invariant and invariant under rigid transformations
generated by $\pphi$ and $\pcsi$, however it is not suitable for fixing
the gauge. The suitable measure is 
$$d[\alpha]=d\pphi d\xi\,,\eqno(3.9)$$
which is gauge invariant and invariant under rigid transformations
generated by $\pphi$ and $\pcsi$.  Furthermore it is expressed in function
of $\xi$. This allows to enforce the gauge fixing procedure. 

In this representation $\{\xi,\pphi\}$ are differential operators. We have
$$ \hat \pcsi\ra -i\d_\xi\,,~~~~\hat\varphi\ra i\d_{\pphi}\,,~~~~~
\hat\xi\ra\xi\,,~~~~~\hat\pphi\ra\pphi\,.\eqno(3.10)$$
Thus the WDW equation is
$$\left(-i\d_\xi-{1\over
16}\pphi^2\right)\Psi(\xi,\pphi)=0\,.\eqno(3.11)$$
The solutions of (3.11) that are eigenstates of $\hat\pphi$ with
eigenvalue $k$ are
$$\Psi_k(\pphi,\xi)=C(k)\delta(\pphi-k)e^{ik^2\xi/16}\,.\eqno(3.12)$$
Now we have to fix the gauge. There is a class of viable gauges for which
there are no Gribov copies and the FP determinant $\de$ is invariant under
gauge transformations. This can be proved as in [5]. Let us simply
choose $\xi$ as time, i.e.\  take
$$\Theta(\xi,\pphi)=\xi-t\eqno(3.13)$$
($t$ is the gauge fixed time parameter); then $\Delta_{FP}=1$. This gauge
is unique and finally the gauge fixed scalar product is
$$(\Psi_2,\Psi_1)\,=\,\int d\pphi \Psi_2^*(\pphi,t) \Psi_1(\pphi,t)\,.
\eqno(3.14)$$
of course a positive definite Hilbert space. Note that the seemingly
obvious choice for the gauge fixing $\Theta'\equiv\gamma - t=0$ (or also
$a-t=0$) leads to the non positive definite scalar product usual in the
Klein-Gordon case ($\Delta'_{FP} = p_{\gamma}$ ); it does not allow a
first quantization interpretation and needs reinterpretation as a second
quantized field. 

The gauge fixed functions in the representation $\{\varphi,\xi=t\}$
read
$$\Psi_k(\varphi,t)={1\over\sqrt{2\pi}}
e^{ik\varphi+ik^2t/16}\,,\eqno(3.15)$$
obviously orthonormal in the Fourier transformed gauge fixed measure
$d\vf$. 

Let us discuss now the reduced method [11]. We impose the gauge 
identity $\xi-t=0$ that gives the effective Hamiltonian
$$H_{\rm eff}=-{1\over 16}\hat\pphi^2\,.\eqno(3.16)$$
The gauge identity implies $\mu=1$ since from the definition of $\xi$ and
the classical general solution of the gauge equations it follows
$\xi=\tau+{\rm const.}$. The \Schr\ equation is 
$$i{\d~\over\d t}\psi(\xi,\pphi)=-{1\over
16}\hat\pphi^2\psi(\xi,\pphi)\,.\eqno(3.17)$$
The stationary eigenfunctions of $\hat\pphi$ coincide with (3.15) and are
orthonormal in the reduced space measure. This proves the equivalence of 
the two quantization procedures.
\bigskip
\leftline{\it b) Quantization in \Sh\ Variables}
\medskip
We can quantize the system also in the \Sh\ representation. Performing the
canonical transformation to the new variables the action becomes
$$S=\int dt\{\dot x\px+\dot y\py-\mu\py\}\,.\eqno(3.18)$$
Let us first quantize the system by the Dirac method. The first step is
the determination of the measure in the inner product (3.8). The
requirement of invariance of the measure under the rigid transformations
generated by $\px$ or $x$ and the gauge transformation generated by $\py$
selects $d[\alpha]=dxdy$ (equivalently $d[\alpha]=d\px dy$), where
$-\infty<x,y,\px<\infty$. The measure $d[\alpha]=d\px d\py$ cannot
be chosen since the gauge fixing function must contain $y$. So, consider
the measure $d\mu=dxdy$: the conjugate variables $\px$ and $\py$ are
represented as
$$\hat \px\ra -i\d_x\,,~~~~\hat\py\ra -i\d_y\,,~~~~~
\hat x\ra x\,,~~~~~\hat y\ra y\,,\eqno(3.19)$$
and the WDW equation becomes
$$-i\d_y\Psi(x,y)=0\,.\eqno(3.20)$$
The solutions of (3.20) that are eigenfunctions of $\hat\px$ with
eigenvalue $k$ are
$$\Psi_k(x)=C(k)e^{ikx}\,.\eqno(3.21)$$
Now we introduce the gauge fixing. The convenient gauge is
$$\Theta(x,y)=y-t\,.\eqno(3.22)$$
Obviously this gauge is unique and $\de=1$. The wave functions (3.21) 
are of course orthonormal (choosing $C(k)=(2\pi)^{-1/2}$) in the inner
product so defined. 

Let us now quantize the system by the alternative method of reducing first
the phase space by a canonical identity. Again the gauge fixing condition
is $y=t$ which determines the Lagrange multiplier as $\mu=1$. Using the
constraint $H=0$ and the gauge fixing condition, the effective Hamiltonian
on the gauge shell becomes $H_{\rm eff}=-\hat\py=0$. The reduced space
Schr\"odinger equation just tells that fixed gauge wave functions do not
depend on $y$. Diagonalizing $\hat\px$ we obtain again the wave functions
(3.21). The two quantization methods give the identical gauge fixed
positive norm Hilbert space. 

We have seen that the quantization of the system can be successfully
completed both in hybrid and \Sh\ variables. The two quantization
procedures are equivalent. Further, the sets of physical wave functions
(3.15) and (3.21) coincide when represented in the same variables. Let us
discuss this point. 

In order to relate the two representations (3.10) and (3.19) we
need the generating function $F$ of the canonical transformation between
the \Sh\ and the hybrid variables: 
$$F(\varphi,\xi;\px,\py)=\varphi\px+\xi\py+{1\over 16}\xi\px^2\,.
\eqno(3.23)$$
The relation between the wave functions in the two representations is
given by
$$\Psi(\xi,\varphi)=\int d\px
d\py~e^{iF(\varphi,\xi;\px,\py)}\Psi(\px,\py)\,.\eqno(3.24)$$
Substituting in (3.24) the Fourier transform of the wave functions (3.21)
$$\Psi(\px,\py)=\delta(\px-k)\delta(\py)\,,\eqno(3.25)$$
it is straightforward to obtain (3.15). This proves the equivalence 
between the hybrid and \Sh\ representation. 

In the \Sh\ variables the reduced Hamiltonian coincides with the original
$H$ and vanishes. The reason is that after the time gauge fixing we are
left with gauge invariant variables; hence inner products and matrix
elements are purely algebraic relations because all operators are built
from classical constant of the motion. The wave functions contain one less
variable because there is no dependence on the gauge fixed time. 

On the contrary, the gauge fixed wave functions for hybrid variables
evolve with time, and the reduced Hamiltonian does not vanish, so these
variables seem to contain more physics.  However the physical content is
the same.  The time dependence expresses the fact that the hybrid
observables are function of time and of the observable gauge invariant
quantities. 

Let us conclude this section noting that since the Hamiltonian (3.1) is
essentially symmetric for
$\{\gamma,\pg\}\leftrightarrow\{\varphi,\pphi\}$, both in the classical
and the quantum treatment one can use the $\{\varphi,\pphi\}$ degrees of
freedom to define the time. 
\beginsection 4. Non Vanishing Cosmological Constant.
This case corresponds to the case treated in [3,4] by GV. The Hamiltonian
is
$${\cal H}=\mu\left[{\pg^2\over 12}-{\pphi^2\over
16}+\Lambda e^{-4\varphi}\right]\,.\eqno(4.1)$$
Again the gauge equations generated by $H$ are integrable. We have
$$\eqalignno{&\gamma=\gamma_0+{\pg\over 6}\tau\,,&\hbox{(4.2a)}\cr
&\pg={\rm constant}\,,&\hbox{(4.2b)}\cr
&e^{2\varphi}=\pm{\sqrt{\Lambda}\over\omega}
\sinh(\omega\tau)\,,&\hbox{(4.2c)}\cr
&\pphi=-4\omega\coth(\omega\tau)\,,&\hbox{(4.2d)}\cr
&\tau=\int_{t_0}^t \mu(t')~dt'\,,~~~~~~\mu(t)>0\,,&\hbox{(4.2e)}\cr}$$
where 
$$\omega=\pm\sqrt{{\pphi^2\over 16}-\Lambda 
e^{-4\varphi}}\,.\eqno(4.3)$$
In (4.2c) the two signs correspond to the sign of $\tau$. $\gamma_0$,
$\pg$ and $\omega$ are gauge invariant. On the constraint $H=0$,
$\pg=\pm\sqrt{12}|\omega|$. The choice of positive $\pg$ corresponds to the
choice of a pre-big bang accelerated expansion $\tau>0$ and a post-big
bang decelerating expansion $\tau<0$ at the basis of the string cosmology
(see e.g.\ [4]). 

Note that $\gamma$ and $\pg$ transform very simply for gauge
transformations; formulae (3.2a,b) hold. This fact will be exploited
later. Let us connect the GV gauge parameter $\tv$ to our
gauge parameter $\tau$. The two parameters are related by 
$$d\tv~=~e^{-2\vf} d\tau\,,\eqno(4.4)$$
that is
$$\sinh(|\omega|\tau)\,\sinh(\tv \sl)=-1\,. \eqno(4.5)$$
The use of $\tau$ is suggested by the simplicity of Eqs.\ (4.2) with the
choice (2.5) of the Lagrange multiplier. From these equations it is easy
to obtain the on-shell solutions of GV [4] that we report for
completeness: 
\medskip
\leftline{$\bullet$ Pre-big bang regime, $\tv<0$:}
$$\eqalign{&a=a_0\left[\tanh\left(-{\tv\sl\over
2}\right)\right]^{-1/\sqrt{3}}\,,\cr
&2(\varphi-\varphi_0)=-\ln\left[\sinh\left(-\tv\sl\right)\right]
\,;\cr}\eqno\hbox{(4.6a)}$$
\medskip
\leftline{$\bullet$ Post-big bang regime, $\tv>0$:}
$$\eqalign{&a=a_0\left[\tanh\left({\tv\sl\over
2}\right)\right]^{1/\sqrt{3}}\,,\cr
&2(\varphi-\varphi_0)=-\ln\left[\sinh\left(\tv\sl\right)\right]
\,.\cr}\eqno\hbox{(4.6b)}$$
\medskip
As in the D'Alembert case, we can define ``hybrid'' and \Sh\ variables.
The hybrid variables are $\{\varphi,\pphi,\xi,\pcsi\}$ defined as in
section 3. The \Sh\ canonical set is $\{w,\pw,z,\pz\}$ defined by
$$\eqalignno{&w\equiv\omega\,,&\hbox{(4.7a)}\cr
&\pw=-12\omega{\gamma\over\pg}-2\,\arcth\left({4\omega\over\pphi}\right)\,,
&\hbox{(4.7b)}\cr
&z\equiv\xi=6{\gamma\over\pg}\,,&\hbox{(4.7c)}\cr
&\pz\equiv H={1\over 12}\pg^2-{1\over 
16}\pphi^2+\Lambda e^{-4\varphi}\,.&\hbox{(4.7d)}\cr}$$
All variables  are gauge invariant except $z$ ($\delta z=\epsilon$); 
$w$ and $\pw$ generate rigid symmetry transformations. Let us quantize 
now the system along the lines of section 3.
\bigskip
\leftline{\it a) Quantization in \Sh\ Variables}
\medskip
Performing the canonical transformation to the \Sh\ variables the action
becomes
$$S=\int dt\{\dot w\pw+\dot z\pz-\mu\pz\}\,.\eqno(4.8)$$
Let us quantize first the system by the Dirac method. The requirement of
invariance of the measure under the rigid transformations generated by $w$
or $\pw$ and the gauge transformation generated by $\pz$ selects the
measures $d[\alpha]=dwdz$ or equivalently $d[\alpha]=d\pw dz$, where
$-\infty<w,z,\pw<\infty$. Given for instance the first one, we have the
representation of the conjugate variables as differential operators: 
$$ \hat \pw\ra -i\d_w\,,~~~~\hat\pz\ra -i\d_z\,,~~~~~
\hat w\ra w\,,~~~~~\hat z\ra z\,.\eqno(4.9)$$
The WDW equation becomes
$$-i\d_z\Psi(w,z)=0\,.\eqno(4.10)$$
The solutions of (4.10) that are eigenfunctions of $\hat w$ with 
eigenvalues $k$ are
$$\Psi_k(w)=C(k)\delta(w-k)\,.\eqno(4.11)$$
The gauge can be fixed as
$$\Theta(w,z)=z-t=0\,.\eqno(4.12)$$
($t$ is thus the fixed gauge time). So the scalar product is defined as
$$(\Psi_2,\Psi_1)\,=\,\int dw\Psi_2^*(w)\Psi_1(w)\,.
\eqno(4.13)$$
Choosing $C(k)=1$, the eigenfunctions (4.11) are orthonormal
in the gauge fixed measure above. 

Let us now quantize the system by the reduced method. Again the gauge
fixing condition is $z=t$. As for the case of section 3, this choice
determines the Lagrange multiplier as $\mu=1$. Using the constraint $H=0$
and the gauge fixing condition, the effective Hamiltonian on the gauge
shell becomes $H_{\rm eff}=-\hat\pz=0$ (typical of the \Sh\ choice of
coordinates). The wave functions do not depend on $z$ and all matrix
elements are of purely algebraic nature. Diagonalizing $\hat w$ we obtain
again the wave functions (4.11) in the reduced Hilbert space. As in the
D'Alembert case, this proves the equivalence of the Dirac and reduced
quantization methods in the representation used. 
\bigskip
\leftline{\it b) Quantization in Hybrid Variables}
\medskip
Let us begin using the Dirac method. As in the case of section 3 we have
to choose the representation and establish the measure. Quite analogously,
the right measure is (3.9). In this case it is better to work in the
Fourier transformed space, so
$$d[\alpha]=d\varphi d\xi\,.\eqno(4.14)$$
Note that (4.14) is not gauge invariant nor invariant under rigid
transformations. However it is related to (3.9) by a Fourier 
transformation.

In the representation $\{\xi,\varphi\}$ the conjugate variables to 
$\xi$ and $\varphi$ are differential operators. We have
$$ \hat \pcsi\ra -i\d_\xi\,,~~~~\hat\pphi\ra -i\d_{\varphi}\,,~~~~~
\hat\xi\ra\xi\,,~~~~~\hat\varphi\ra\varphi\,.\eqno(4.15)$$
The WDW equation is
$$\left(-i\d_\xi+{1\over 16}\d_\varphi^2+\Lambda 
e^{-4\varphi}\right)\Psi(\xi,\varphi)=0\,.\eqno(4.16)$$
The solutions of (4.16) that are eigenstates of $\hat \omega$
with eigenvalue $k$ are of the form
$$\Psi_k(\varphi,\xi)=\left[A_{+}(k)Z_{2ik}\left(2\sqrt{\Lambda}
e^{-2\varphi}\right)+A_{-}(k)Z_{-2ik}\left(2\sqrt{\Lambda}
e^{-2\varphi}\right)\right]e^{ik^2\xi}\,,\eqno(4.17)$$
where $Z$ is a generic linear combination of Bessel functions.
The choice
$$\Psi_k(\varphi,\xi)=A(k)J_{2ik}e^{ik^2\xi}\eqno(4.18)$$
has been selected by GV as representing the reflection of the WDW wave
function correspondent to the birth of a decelerating expanding universe. 

Now we have to fix the gauge. Using $\Theta=\xi-t$
the definition of the inner product is
$$(\Psi_2,\Psi_1)\,=\,\int d\vf  \Psi^*_2(\vf,t) \Psi_1(\vf,t)\,=
\,\int {{dz}\over{z}} \Psi^*_2(z,t)\Psi_1(z,t)\,.
\eqno(4.19)$$
where $z=2\sl e^{2\vf}$. Note that the choice $\gamma=t$ does not yield a
positive definite norm. 

The two sets of real orthonormal functions in the gauge
fixed measure (4.19): 
$$\eqalignno{&\chi^{(1)}_k(z,t)=\sqrt{k\cosh(\pi k)\over 2\sinh(\pi 
k)}\left[e^{-\pi k}H^{(1)}_{2ik}(z)+e^{\pi 
k}H^{(2)}_{2ik}(z)\right]e^{ik^2t}\,,&\hbox{(4.20a)}\cr\cr
&\chi^{(2)}_k(z,t)=i\sqrt{k\sinh(\pi k)\over 2\cosh(\pi k)}
\left[e^{-\pi k}H^{(1)}_{2ik}(z)-e^{\pi 
k}H^{(2)}_{2ik}(z)\right]e^{ik^2t}\,.&\hbox{(4.20b)}\cr}$$
Let us discuss now the reduced method. The gauge $\xi=t$ gives the
effective Hamiltonian 
$$H_{\rm eff}=-\hat w^2=-{\hat\pphi^2\over 16}+\Lambda
e^{-2\varphi}\,,\eqno(4.21)$$
and the \Schr\ equation coincides with (4.16). The stationary \Schr\
equation is 
$$\left[{1\over 16}\d_\varphi^2+\Lambda e^{-4\varphi}\right] 
\Psi(\varphi)=E\Psi(\varphi)\,,~~~~~~~~~E<0\,.\eqno(4.22)$$
and its solutions are those of Eq.\ (4.17) where $k=\sqrt{-E}$ and
they can be chosen orthonormal as in (4.20). 
\beginack
We thank Maurizio Gasperini and Gabriele Veneziano for interesting 
discussions.
\beginref
\ref{1}{A.\ Vilenkin, talk given at the {\tscors International School of
Astrophysics D.\ Chalonge; 4th Course:  String Gravity and Physics at the
Planck Energy Scale (A NATO Advanced Study Institute)}, Erice, Italy 8-19
Sept.\ 1995, e-Print Archive: gr-qc/9507018 and references therein}

\ref{2}{G.\ Veneziano, \PLB{265}{287}{1991}; M.\ Gasperini and G.\ 
Veneziano, \APP{1}{317}{1993}}

\ref{3}{M.\ Gasperini, J.\ Maharana and G.\ Veneziano, ``Graceful Exit in
Quantum String Cosmology'', CERN Preprint CERN-TH/96-32, e-Print 
Archive: hep-th/9602087}

\ref{4}{M.\ Gasperini and G.\ Veneziano, ``Birth of the Universe as Quantum
Scattering in String Cosmology'', CERN Preprint CERN-TH/96-49,
e-Print Archive: hep-th/9602096}

\ref{5}{For a discussion of the importance that the symmetries of the
classical action have for the definition of the Hilbert space in
minisuperspace models, see for instance: M.\ Cavagli\`a, V.\ de Alfaro,
and A.T.\ Filippov, ``Quantization of the \Sc\ Black Hole'', preprint
DFTT 50/95, August 1995, e-Print Archive: gr-qc/95\-08\-062, accepted for
publication in {\tscors Int.\ Jou.\ Mod.\ Phys.} {\bf D}} 

\ref{6}{For a review of the quantization of gauge systems, see M.\
Henneaux and C.\ Teitelboim, {\it Quantization of Gauge Systems}
(Princeton, New Jersey, 1992)}

\ref{7}{C.\ Lovelace, \PLB{135}{75}{1984}; E.S.\ Fradkin and A.A.\
Tseytlin, \NPB{261}{1}{1985}; C.G.\ Callan, E.J.\ Martinec, M.J.\ Perry,
and D.\ Friedan, \NPB{262}{593}{1985}}

\ref{8}{L.D.\ Landau and E.M.\ Lifshitz, {\it The Classical Theory of
Fields} (Pergamon Press, Oxford, 1975)}

\ref{9}{S.\ Shanmugadhasan, \JMP{14}{677}{1973}}

\ref{10}{C.\ Teitelboim, in: ``Quantum Cosmology and Baby Universes'',
Proceedings of the {\it 7th Winter School for Theoretical Physics},
Jerusalem, Israel, December 27 1989 - January 4, 1990, Ed.\ S.\ Coleman,
J.B.\ Hartle, T.\ Piran, and S.\ Weinberg (World Scientific, Singapore,
1991)}

\ref{11}{M. Cavagli\`a, V. de Alfaro and A.T. Filippov,
\IJMPA{10}{611}{1995}}

\endref
\bye